\documentclass[sigconf,nonacm]{acmart}
\usepackage{blindtext,epsfig,endnotes,fancybox}

\renewcommand\footnotetextcopyrightpermission[1]{} 
\setcopyright{none}
\newif\ifcameraready
\camerareadyfalse

\usepackage{amsmath,amsfonts}
\usepackage[T1]{fontenc}
\usepackage[utf8]{inputenc}

\usepackage{upquote}

\usepackage{microtype}
\UseMicrotypeSet[protrusion]{basicmath} 

\usepackage{hyperref}
\usepackage{graphicx,grffile}

\usepackage{xspace}
\usepackage{tabularx}
\usepackage{ragged2e}
\usepackage{booktabs}
\usepackage{paralist}
\usepackage[american]{babel}
\usepackage[shortlabels]{enumitem}
\usepackage{courier,color,wrapfig}
\usepackage{colortbl}
\usepackage{xspace}
\usepackage{balance}
\usepackage{enumitem}
\usepackage{epstopdf}
\usepackage{multirow}
\usepackage{booktabs}
\usepackage{url}
\usepackage{pifont}
\usepackage{subcaption}
\usepackage{tikz}
\usepackage{mathtools}
\usepackage[normalem]{ulem}
\usepackage{xkeyval}
\usepackage{comment}

\usepackage[]{algorithm2e}
\RestyleAlgo{boxruled}
\usepackage{etoolbox}
\usepackage[margin=5pt,font=small,labelfont={bf,rm}]{caption}
\usepackage{ifthen}
\usepackage{sepfootnotes}
\usepackage[compact]{titlesec}

\usepackage{listings}

\usetikzlibrary{shapes.geometric}
\usetikzlibrary{shapes.multipart}
\usepackage{pgfplots}
\usepgfplotslibrary{fillbetween}
\usetikzlibrary{pgfplots.groupplots}
\pgfplotsset{compat=1.15}

\lstdefinestyle{cpp}{
  language=[11]C++,
  breaklines=true,
  showstringspaces=false,
  basicstyle={\scriptsize\ttfamily\bfseries},
  commentstyle = {\itshape\bfseries\color{brown}},
  identifierstyle=\color{black},
  stringstyle=\color{black},
  morekeywords = {ntohs,htons},
  keywordstyle = {\bfseries\color{blue!60!black}},
  emph = [1]{
    int,
    uint8_t,
    uint16_t,
    uint32_t,
    uint64_t,
    proto_ptr_t,
    Destination_Unreachable_Message_hdr,
    Echo_or_Echo_Reply_Message_hdr,
    Information_Request_or_Information_Reply_Message_hdr,
    Parameter_Problem_Message_hdr,
    Redirect_Message_hdr,
    Source_Quench_Message_hdr,
    Time_Exceeded_Message_hdr,
    Timestamp_or_Timestamp_Reply_Message_hdr,
  },
  emphstyle = [1]{\bfseries\color{blue}},
}

\lstdefinestyle{cpp-dump}{
  style=cpp,
  basicstyle={\tiny\ttfamily\bfseries},
}

\DeclareMathDelimiter{(}{\mathopen} {operators}{"28}{largesymbols}{"00}
\DeclareMathDelimiter{)}{\mathclose}{operators}{"29}{largesymbols}{"01}

\PassOptionsToPackage{usenames,dvipsnames}{color} 
\hypersetup{unicode=true,
            colorlinks=true,
            linkcolor=blue,
            citecolor=blue,
            anchorcolor=blue,
            urlcolor=blue,
            breaklinks=true}
\makeatletter
\def\maxwidth{\ifdim\Gin@nat@width>\linewidth\linewidth\else\Gin@nat@width\fi}
\def\maxheight{\ifdim\Gin@nat@height>\textheight\textheight\else\Gin@nat@height\fi}
\makeatother
\setkeys{Gin}{width=\maxwidth,height=\maxheight,keepaspectratio}
\makeatletter
\g@addto@macro{\UrlBreaks}{\UrlOrds}
\makeatother

\apptocmd\normalsize{%
\abovedisplayskip=5pt
\abovedisplayshortskip=5pt
\belowdisplayskip=5pt
\belowdisplayshortskip=5pt
}{}{}

\makeatletter
\renewcommand{\verbatim@font}{\ttfamily\bfseries\footnotesize}
\makeatother

\definecolor{AccentedRowColor}{rgb}{0.95, 1, 0.65}
\newcommand*\rot{\rotatebox{60}}
\newcommand*\OK{{\sffamily x}}
\newcommand*\fullmark{\textcolor{green}{$\blacklozenge$}\xspace}
\newcommand*\partialmark{\textcolor{olive}{$\boldsymbol{+}$}\xspace}

\newcommand{\sysname}{\textsc{sage}\xspace}
\newcommand{\system}{\textsc{sage}\xspace}

\newcommand\paraspace{\vspace*{1ex}}
\newcommand\parab[1]{\paraspace\noindent\textbf{#1}}
\newcommand\parae[1]{\textbf{\textit{#1}}}

\newcommand\tableh[1]{\textsc{\textbf{#1}}}

\newcommand{\etc}{\emph{etc.}\xspace}
\newcommand{\ie}{\emph{i.e.,}\xspace}
\newcommand{\eg}{\emph{e.g.,}\xspace}

\newcommand{\secref}[1]{\S\ref{#1}}
\newcommand{\figref}[1]{Figure~\ref{#1}}
\newcommand{\tabref}[1]{Table~\ref{#1}}

\def\predicate#1{\textsc{#1}}

\begin{document}

\title{\huge Semi-Automated Protocol Disambiguation and Code Generation}

\author{Jane Yen}
\affiliation{%
  \institution{University of Southern California}
}
\email{yeny@usc.edu}

\author{Tamás Lévai}
\affiliation{%
  \institution{Budapest University of Technology and Economics}
}
\email{levait@tmit.bme.hu}

\author{Qinyuan Ye}
\affiliation{%
  \institution{University of Southern California}
}
\email{qinyuany@usc.edu}

\author{Xiang Ren}
\affiliation{%
  \institution{University of Southern California}
}
\email{xiangren@usc.edu}

\author{Ramesh Govindan}
\affiliation{%
  \institution{University of Southern California}
}
\email{ramesh@usc.edu}

\author{Barath Raghavan}
\affiliation{%
  \institution{University of Southern California}
}
\email{barathra@usc.edu}


\begin{abstract}
For decades, Internet protocols have been specified using natural language. Given the ambiguity inherent in such text, it is not surprising that protocol implementations have long exhibited bugs. In this paper, we apply natural language processing (NLP) to effect semi-automated generation of protocol implementations from specification text. Our system, \sysname, can uncover ambiguous or under-specified sentences in specifications; once these are clarified by the spec author, \sysname can generate protocol code automatically.

Using \sysname, we discover 5 instances of ambiguity and 6 instances of under-specification in the ICMP RFC; after clarification, \sysname is able to automatically generate code that interoperates perfectly with Linux implementations. We show that \sysname generalizes to BFD, IGMP, and NTP. We also find that \sysname supports many of the conceptual components found in key protocols, suggesting that, with some additional machinery, \sysname may be able to generalize to TCP and BGP. 






\end{abstract}

\maketitle
\thispagestyle{empty}


\section{Introduction}
\label{s:introduction}

Four decades of Internet protocols have been specified in English and used to create, in Clark's words, rough consensus and running code~\cite{clark-saying}.  In that time we have come to depend far more on network protocols than most imagined. To this day, engineers implement a protocol by reading and interpreting specifications as described in Request For Comments documents (RFCs).  Their challenge is to navigate easy-to-misinterpret colloquial language while writing not only a bug-free implementation but also one that \emph{interoperates} with code written by another person at a different time and place.


Software engineers find it difficult to interpret specifications in large part because natural language can be ambiguous. Unfortunately, such ambiguity is not rare; the errata alone for RFCs over the years highlight numerous ambiguities and the problems they have caused~\cite{rfc8448,rfc8452, rfc8492, rfc8519}. Ambiguity has resulted in buggy implementations, security vulnerabilities, and expensive and time-consuming software engineering processes, like interoperability bake-offs~\cite{sip_bakeoff,ipp_bakeoff}.

To address this, one line of research has sought formal specification of programs and protocols (\secref{s:related-work}), which would enable verifying spec correctness and, potentially, enable automated code generation~\cite{boussinot1991esterel}.  However, formal specs are cumbersome and thus have not been adopted in practice; to date, protocols are specified in natural language.\footnote{In recent years, attempts have been made to formalize other aspects of network operation, such as network configuration~\cite{RPSL,Propane} and control plane behavior~\cite{Pyretic}, with varying degrees of success.}


In this paper, we apply NLP to semi-automated generation of protocol implementations from RFCs. 
Our main challenge is to understand the \textit{semantics} of a specification. This task, \textit{semantic parsing}, has advanced in recent years with parsing tools such as CCG~\cite{artzi2013semantic}. Such tools describe natural language with a \textit{lexicon} and yield a semantic interpretation for each sentence. Because they are trained on generic prose, they cannot be expected to work out of the box for idiomatic network protocol specifications, which contain embedded syntactic cues (\eg structured descriptions of fields), incomplete sentences, and implicit context from neighboring text or other protocols. More importantly, the richness of natural language will likely always lead to ambiguity, so we do not expect fully-automated NLP-based systems (\secref{sec:background}).



\parab{Contributions.} In this paper, we describe \sysname, a \textit{semi-automated} approach to protocol generation from natural language. \sysname is iterative: it reads RFC text and marks sentences (a) for which it cannot generate unique semantic interpretations or (b) which fail on the protocol's \textit{unit tests} (\sysname uses test-driven development). The former sentences are likely semantically ambiguous whereas the latter represent under-specified behaviors. In either case, the spec author can then revise the sentences (perhaps several times), until the resulting RFC can cleanly be turned into code.

We make the following contributions.
First, we tailor (\secref{section:approach}) a pre-existing semantic parser to extend its vocabulary and lexicon to cover networking-specific language in RFCs. Our extensions also extract structural context from RFCs that may be crucial to understand a sentence's semantics and to generate code. Even so, such a parser can fail to extract a sentence's semantics or emit \textit{multiple} semantic interpretations; ideally, the parser would generate exactly one interpretation.

Second, we note that many challenges stem from RFC idioms, such as incomplete sentences to describe protocol header fields and the use of verbs like \textit{is} and prepositions like \textit{of} in specific ways. Using these conventions, we devise automated techniques to eliminate many semantic interpretations. If, after such \textit{disambiguation} (\secref{section:disambiguation}), a sentence has multiple interpretations, it is likely to be fundamentally ambiguous, so \sysname requires a human to rewrite the sentence.

\begin{table}[t]
  \centering
  \footnotesize
  \begin{tabular}{l l}
    \toprule
    \tableh{Name} & \tableh{Description}\\
    \midrule
    \textbf{\fullmark Packet Format} & Packet anatomy (i.e., field structure) \\
    \textbf{\fullmark Field Descriptions} & Packet header field descriptions \\
    \textbf{\fullmark Constraints} & Constraints on field values \\
    \textbf{\fullmark Protocol Behaviors} & Reactions to external/internal events \\
    \textbf{System Architecture} & Protocol implementation components \\
    \textbf{\partialmark State Management} & Session information and/or status \\
    \textbf{Comm. Patterns} & Message sequences (e.g., handshakes) \\
    \bottomrule
  \end{tabular}
  \caption{Protocol specification components. \sysname supports those marked with \fullmark (fully) and \partialmark (partially).}
  \label{tab:brief-protocol-component}
  \vskip -1em
\end{table}

Third, we present a \textit{code generator} that converts semantic representations to executable code (\secref{section:codegen}). The parser represents a sentence's semantics using logical predicates, and the code generator emits executable code using contextual information that it has gleaned from the RFC's document structure, as well as static context pre-defined in \sysname about lower-layer protocols and the underlying OS. Unit testing on generated code can uncover incompleteness in specifications.

\sysname discovered (\secref{section:evaluation}) 5 sentences in the ICMP RFC~\cite{rfc792} (of which 3 are unique, the others being variants) that had multiple semantic interpretations even after disambiguation. It also discovered 6 sentences that failed unit tests (all variants of a single sentence). After fixing these, \sysname was able to automatically generate code for ICMP that interoperated perfectly with \verb=ping= and \verb=traceroute=. In contrast, graduate students asked to implement ICMP in a networking course made numerous errors (\secref{sec:background}). Moreover, \sysname was able to parse significant sections of BFD~\cite{rfc5880}, IGMP~\cite{rfc1112}, and NTP~\cite{rfc1059}, with few additions to the lexicon. It generated packets for the timeout procedure containing both NTP and UDP headers. It also parsed state management text for BFD to determine system actions and update state variables for reception of control packets. Finally, \sysname's disambiguation is often very effective, reducing, in some cases, 56 logical forms to 1.


\parab{Toward greater generality.} 
\sysname is a significant first step toward automated processing of natural-language protocol specifications, but much work remains. Protocol specifications contain a variety of components, and ~\tabref{tab:brief-protocol-component} indicates which ones \sysname supports well (in green), which it supports partially (in olive), and which it does not support.
Some protocols contain complex state machine descriptions (\eg TCP) or describe how to process and update state (\eg BGP); \sysname can parse state management in a simpler protocol like BFD.
Other protocols describe software architectures  (\eg OSPF, RTP) and communication patterns (\eg BGP); \sysname must be extended to parse these descriptions. In \secref{section:discussion}, we break down the prevalence of protocol components by RFC to contextualize our contributions, and identify future \sysname extensions. Such extensions will put \sysname within reach of parsing large parts of TCP and BGP RFCs.

\section{Background and Overview}
\label{sec:background}

Spec ambiguities can lead to bugs and non-interoperability, which we quantify next  via implementations of ICMP~\cite{rfc792} by students in a graduate networking course.


\subsection{Analysis of ICMP Implementations}
\label{s:analys-impl}

ICMP, defined in RFC 792 in 1981 and used by core tools like \verb=ping= and \verb=traceroute=, is a simple protocol whose specification should be easy to interpret. To test this assertion, we examined implementations of ICMP by 39 students in a graduate networking class. Given the ICMP RFC and related RFCs, students built ICMP message handling for a router.

To test whether students implemented echo reply correctly, we used the Linux \verb=ping= tool to send an echo message to their router (we tested their code using Mininet~\cite{lantz2010network}). Across the 39 implementations, the Linux implementation correctly parsed the echo reply only for 24 of them (61.5\%). One failed to compile and the remaining 14 exhibited 6 categories (not mutually exclusive) of implementation errors (\tabref{t:error_type}): mistakes in IP or ICMP header operations; byte order conversion errors; incorrectly-generated ICMP payload in the echo reply message; incorrect length for the payload; and wrongly-computed ICMP checksum. Each error category occurred in at least 4 of the 14 erroneous implementations.


\begin{table}
  \centering
  \footnotesize
  \begin{tabular}{lr}
    \toprule
    \tableh{Error Type} & \tableh{Frequency} \\
    \midrule
    IP header related & 57\% \\ 
    ICMP header related & 57\% \\ 
    Network byte order and host byte order conversion & 29\% \\ 
    Incorrect ICMP payload content & 43\% \\ 
    Incorrect echo reply packet length & 29\% \\ 
    Incorrect checksum or dropped by kernel & 36\% \\ 
    \bottomrule
  \end{tabular}
  \caption{Error types of failed cases and their frequency in 14 faulty student ICMP implementations.}
  \label{t:error_type}
  \vskip -1.5em
\end{table}

To understand the incorrect checksum better, consider the specification of the ICMP checksum in this sentence: \textit{The checksum is the 16-bit one's complement of the one's complement sum of the ICMP message starting with the ICMP Type.}
This sentence does not specify where the checksum should \textit{end}, resulting in a potential ambiguity for the echo reply; a developer could checksum some or all of the header, or both the header and the payload. In fact, students came up with \textit{seven} different interpretations (\tabref{t:cksum_variety}) including checksumming only the IP header, checksumming the ICMP header together with a few fixed extra bytes, and so on.


\begin{table}
  \centering
  \footnotesize
  \begin{tabular}{cp{6.8cm}}
    \toprule
    \tableh{Index} & \shortstack[c]{\tableh{ICMP checksum range interpretations}} \\ 
    \midrule
     1 &  Size of a specific type of ICMP header. \\ 
     2 & Size of a partial ICMP header. \\ 
     3 &  Size of the ICMP header and payload. \\ 
     4 & Size of the IP header. \\ 
     5 & Size of the ICMP header and payload, and any IP options. \\ 
     6 & Incremental update of the checksum field using whichever checksum range the sender packet chose.\\ 
     7 & Magic constants (e.g., 2 or 8 or 36).\\ 
     \bottomrule
  \end{tabular}
  \caption{Students' ICMP checksum range interpretations.}
  \label{t:cksum_variety}
\end{table}




\subsection{Approach and Overview}
\label{s:problem-scope}
\vskip -0.5em
\parab{Dealing with Ambiguity.}
These results indicate that even a relatively simple RFC results in a large variety of interpretations and often result in non-interoperable implementations. RFC authors and the IETF community rely on manual methods to avoid or eliminate non-interoperabilities: careful review of standards drafts by participants, development of reference implementations, and interoperability \emph{bake-offs}~\cite{sip_bakeoff,ipp_bakeoff} at which vendors and developers test their implementations against each other to discover issues that often arise from incomplete or ambiguous specs.

\parab{Why not reference implementations?}
Reference implementations are useful but insufficient. For a reference protocol document to become a standard, a reference implementation is indeed often written, and this has been the case for many years. A reference implementation is often written by participants in the standardization process, who may or may not realize that there exist subtle ambiguities in the text. Meanwhile, vendors write code directly to the specification (often to ensure that the resulting code has no intellectual property encumbrances), sometimes many years after the spec was standardized. This results in subtle incompatibilities in implementations of widely deployed protocols~\cite{pedrosa2015analyzing}.

\parab{Why are there ambiguities in RFCs?}
RFCs are ambiguous because (a) natural language is expressive and admits multiple ways to express a single idea; (b) standards authors 
are technical domain experts who may not always recognize the nuances of natural language; and (c) context matters in textual descriptions, and RFCs may omit context.

\parab{Structure in RFCs.}
Unlike general English text, network protocol specifications have exploitable structure. The networking community uses a restricted set of words and operations (\ie \textit{domain-specific} terminology) to describe network behaviors. Moreover, RFCs conform to a uniform style~\cite{rfc7322} (especially recent RFCs) and all standards-track RFCs are carefully edited for clarity and style adherence~\cite{rfced}.

\parab{Our approach.} We leverage recent advances in the NLP area of \textit{semantic parsing}. Natural language can have lexical~\cite{rayner1986lexical, kempson1981ambiguity} (\eg the word \textit{bat} can have many meanings), structural (\eg the sentence \textit{Alice saw Bob with binoculars}) and semantic (\eg in the sentence \textit{I saw her duck}) ambiguity.

For the foreseeable future we do not expect NLP to be able to parse RFCs without help. Thus, our \emph{semi-automated} approach uses NLP tools, along with unit tests, to help a human-in-the-loop discover and correct ambiguities after which the spec is amenable to automated code generation.

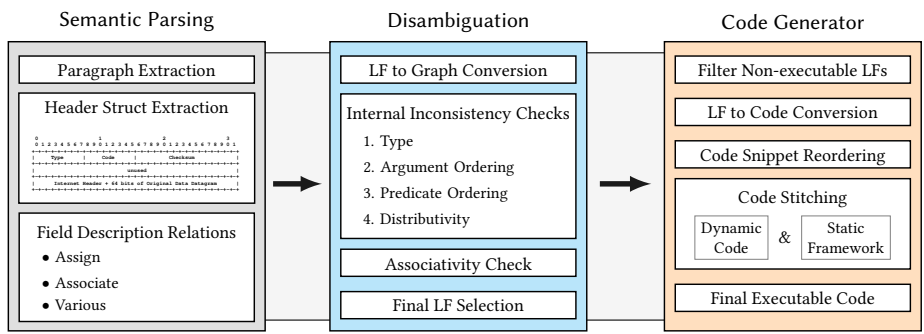
\begin{figure*}
    \centering
    \resizebox{.7\textwidth}{!}{\tikzset{%
  box/.style = {
    draw,
    thick,
    align = center,
    inner sep = 6pt,
    text centered,
  },
  arrow/.style = {
    ->,
    line width = 3,
    white!10!black,
  },
  blocktitle/.style = {
    align = center,
    font = \Large\sffamily,
    yshift = 62.5,
  },
  block/.style = {
    box,
    minimum height = 160,
    minimum width = 142,
    yshift= 20,
    fill = white!80!black,
  },
  component/.style = {
    box,
    fill = white,
    minimum width = 130,
    text centered,
  },
  subcomponent/.style = {
    box,
    thin,
    gray,
    text = black,
    inner sep = 3.5pt,
    font = \small,
    minimum width = 25,
    minimum height = 15,
  },
  listitem/.style = {
    align = left,
    inner sep = 0pt,
    node distance = 0pt,
    text width = 105,
    font = \small,
  },
}

\def\bluedoc {
  \draw [thick,fill=blue, fill opacity=.2]
   (0,0) -- (0, 1) -- (1.5, 1) -- (1.5, 0.25) -- (1.35, 0) -- (0, 0);
  \draw [fill=blue, fill opacity=.4]
   (1.5, 0.25) -- (1.375, 0.2) -- (1.35, 0);
}

\def\indentdoc {
  \bluedoc;
  \node[] at (.75, .58) {
    \scalebox{.15}{
      \lstinputlisting[basicstyle=\fontsize{6}{6}\selectfont\ttfamily\bfseries,xleftmargin=.5em]
      {figs/random_rfc_text.txt}} };
  \node[text=red] at (.112,.81) { \scalebox{.4}{\tiny $\Longleftrightarrow$ } };
  \node[text=red] at (.085,.48) { \scalebox{.55}{\tiny $\Leftrightarrow$ } };
  \node[text=red] at (.185,.27) { \scalebox{.75}{\tiny $\Longleftrightarrow$ } };
}

\def\sectiondoc {
  \bluedoc;
  \node[] at (.75, .58) {
    \scalebox{.15}{
      \lstinputlisting[basicstyle=\fontsize{6}{6}\selectfont\ttfamily\bfseries,xleftmargin=.5em]
      {figs/random_rfc_text.txt}} };
  \draw[red] (.1, .92) rectangle (1.4, .56);
  \draw[red] (.1, .51) rectangle (1.4, .24);
}

\begin{tikzpicture}[>=latex]

  \draw[box, white!25!black, fill=white!96!black] (5, 2.1) rectangle (15, 7.3);

  \draw[arrow] (6.4, 4.75) -- (7.35, 4.75);
  \draw[arrow] (12.75, 4.75) -- (13.75, 4.75);

  \node[block, fill=white!75!gray] at (3.75, 4) (PreProcBlock) {};
  \node[blocktitle, above of=PreProcBlock] {Semantic Parsing};

  \node[component, anchor = north,
  minimum height = 15,
  label = {[yshift=-17.5]Paragraph Extraction}] at (3.75, 7.25) (ParaExtComp) {};

  \node[component, below of = ParaExtComp,
  yshift = -15,
  minimum height = 60,
  label = {[yshift=-15]Header Struct Extraction}] (HeaderComp) {
    \scalebox{.49}{
      \lstinputlisting[basicstyle=\fontsize{6}{6}\selectfont\ttfamily\bfseries,xleftmargin=.5em]
      {figs/ascii_art2.txt}}
  };

  \node[component, below of = HeaderComp,
  yshift = -37.5,
  minimum height = 60,
  label = {[yshift=-20]Field Description Relations}] (FieldDescComp) {};

  \matrix[below of = FieldDescComp, row sep = 6pt, yshift = 21] {
    \node[listitem] {$\bullet$ Assign}; \\
    \node[listitem] {$\bullet$ Associate}; \\
    \node[listitem] {$\bullet$ Various}; \\
  };

  \node[block, fill=white!75!cyan] at (10, 4) (DisambBlock) {};
  \node[blocktitle, above of = DisambBlock] {Disambiguation};

  \node[component, anchor = north,
  minimum height = 15,
  label = {[yshift=-17.5]LF to Graph Conversion}] at (10, 7.25) (LFtoGraphComp) {};

  \node[component, below of = LFtoGraphComp,
   yshift = -25,
   minimum height = 80,
   label = {[yshift=-20]Internal Inconsistency Checks}] (IntInconCheckComp) {};

  \matrix[below of = IntInconCheckComp, row sep = 6pt, yshift = 20] {
    \node[listitem] {$1.$ Type}; \\
    \node[listitem] {$2.$ Argument Ordering}; \\
    \node[listitem] {$3.$ Predicate Ordering}; \\
    \node[listitem] {$4.$ Distributivity}; \\
  };

  \node[component, below of = IntInconCheckComp,
  yshift = -25,
  minimum height = 15,
  label = {[yshift=-17.5]Associativity Check}] (GraphIsoComp) {};

  \node[component, below of = GraphIsoComp,
  yshift = 5,
  minimum height = 15,
  label = {[yshift=-15]Final LF Selection}] (FinLFSelComp) {};

  \node[block, fill = white!75!orange] at (16.5, 4) (CodeGenBlock) {};
  \node[blocktitle, above of=CodeGenBlock] {Code Generator};

  \node[component, anchor = north,
  minimum height = 15,
  label = {[yshift=-15]Filter Non-executable LFs}] at (16.5, 7.25) (LFFilterComp) {};

  \node[component, below of = LFFilterComp,
  yshift = 5,
  minimum height = 15,
  label = {[yshift=-15]LF to Code Conversion}] (LFtoCodeComp) {};

  \node[component, below of = LFtoCodeComp,
  yshift = 5,
  minimum height = 15,
  label = {[yshift=-17.5]Code Snippet Reordering}] (CSReOComp) {};

  \node[component, below of = CSReOComp,
   yshift = -10,
   minimum height = 50,
   label = {[yshift=-20]Code Stitching}] (CodeStitchComp) {};

  \matrix[below left of = CodeStitchComp, xshift=20, yshift = 12.5, column sep = 2.25] {
    \node[subcomponent] (DynCodeSComp) {Dynamic\\ Code}; &
    \node[] {\&}; &
    \node[subcomponent] (StatCodeSComp) {Static\\ Framework}; \\
  };


  \node[component, below of = CodeStitchComp,
  yshift = -12.5,
  minimum height = 15,
  label = {[yshift=-15]Final Executable Code}] (FinalCodeComp) {};

\end{tikzpicture}

}
    \caption{SAGE components.} 
    \label{fig:sys_pipeline}
    \vskip -1em
\end{figure*}

\figref{fig:sys_pipeline} shows the three stages of \sysname. The \textit{parsing} stage uses a semantic parser~\cite{artzi2013semantic} to generate intermediate representations, called \textit{logical forms} (LFs), of sentences. Because parsing is not perfect, it can output multiple LFs for a sentence. Each LF corresponds to one semantic interpretation of the sentence, so multiple LFs represent ambiguity. The \textit{disambiguation} stage aims to automatically eliminate such ambiguities. If, after this, ambiguities remain, \sysname asks a human to resolve them. The \textit{code generator} compiles LFs into executable code, a process that may also uncover ambiguity. 

\section{Semantic Parsing}
\label{section:approach}

\textit{Semantic parsing} is the task of extracting meaning from a document.  Tools for semantic parsing formally specify natural language grammars and extract parse trees from text. More recently, deep-learning based approaches have proved effective in semantic parsing~\cite{dong2018coarse, yin2018structvae, krishnamurthy2017neural} and certain types of automatic code generation~\cite{yin2017syntactic,lin2017program,rabinovich2017abstract}.
However, such methods do not directly apply to our task. First, deep learning typically requires training in a ``black-box''. Since we aim to identify ambiguity in specifications, we aim to interpret intermediate steps in the parsing process and maintain all valid parsings. Second, such methods require large-scale annotated datasets; collecting high-quality data that maps network protocol specifications to expert-annotated logical forms (for supervised learning) is impractical. 

For these reasons, we use the Combinatory Categorial Grammar (CCG~\cite{artzi2013semantic}) formalism that enables (a) coupling syntax and semantics in the parsing process and (b) is well suited to handling domain-specific terminology by defining a small hand-crafted lexicon that encapsulates domain knowledge. CCG has successfully parsed natural language explanations into labeling rules in several contexts~\cite{srivastava2017joint, Wang2020Learning}.


\parab{CCG background.}
\label{sec:ccg-background}
CCG takes as input a description of the language syntax and semantics. It describes the syntax of words and phrases using \textit{primitive categories} such as noun (N), noun phrase (NP), or sentence (S), and \textit{complex categories} comprised of primitive categories, such as S$\backslash$NP (to express that it can combine a noun phrase on the left and form a sentence). It describes semantics with lambda expressions such as $\lambda x. \lambda y. \predicate{@Is}(y,x)$ and $\lambda x. \predicate{@Compute}(x)$.

CCG encodes these in a \textit{lexicon}, which users can extend to capture domain-specific knowledge. For example, we added the following lexical entries to the lexicon to represent constructs found in networking standards documents:
{
\setitemize{nolistsep}
\begin{enumerate}[topsep=1ex]
    \setlength\itemsep{0em}
    \item checksum $\rightarrow$ {NP: "checksum"}
    \item is $\rightarrow$ \{(S{\textbackslash}NP)/NP: $\lambda x. \lambda y. \predicate{@Is}(y,x)$\}
    \item zero $\rightarrow$ \{NP: \predicate{@Num}(0)\}
\end{enumerate}
}
This expresses the fact (a) ``checksum'' is a special word in networking, (b) ``is'' can be assignment, and (c) zero can be a number.
CCG can use this lexicon to generate a \textit{logical form} (LF) that completely captures the semantics of a phrase such as ``checksum is zero'': \{S: $\textsc{@Is}(\text{"checksum"},\textsc{@Num}(0))$\}. Our code generator (\secref{section:codegen}) produces code from these.

\parab{Challenges.} \sysname must surmount three challenges before using CCG: (a) specify \textit{domain-specific} syntax, (b) specify \textit{domain-specific} semantics, (c) extract \textit{structural} and \textit{non-textual} elements in standards documents (described below). Next we describe how we address these challenges.

\parab{Specifying domain-specific syntax.} Lexical entry (1) above specifies that \emph{checksum} is a keyword in the vocabulary. Rather than having a person specify such syntactic lexical entries, \sysname creates a \textit{term dictionary} of domain-specific nouns and noun-phrases using the index of a standard networking textbook. This reduces human effort. Before we run the semantic parser, CCG, we also need to identify nouns and noun-phrases that occur generally in English, for which we use an NLP tool called SpaCy~\cite{spacy2}.

\parab{Specifying domain-specific semantics.} CCG has a built-in lexicon that captures the semantics of written English. Even so, we have found it important to add domain-specific lexical entries. For example, the lexical entry (2) above shows that the verb \textit{is} can represent the assignment of a value to a protocol field. In \sysname, we manually generate these domain-specific entries, with the intent that these semantics will generalize to many RFCs (see also \secref{section:evaluation}). Beyond capturing domain-specific uses of words (like \textit{is}), domain-specific semantics capture idiomatic usage common to RFCs. For example, RFCs have field descriptions (like version numbers, packet types) that are often followed by a single sentence that has the (fixed) value of the field. For CCG to parse this, it must know that the value should be assigned to the field. Similarly, RFCs sometimes represent descriptions for different code values of a type field using an idiom of the form \mbox{``0 = Echo Reply''}. \secref{section:evaluation} quantifies the work involved in generating the domain-specific lexicon.

\parab{Extracting structural and non-textual elements.} Finally, RFCs contain stylized elements, for which we wrote pre-processors. RFCs use indentation to represent content hierarchy and descriptive lists (\eg field names and their values). Our pre-processor extracts these relationships to aid in disambiguation (\secref{section:disambiguation}) and code generation (\secref{section:codegen}). RFCs also represent packet header fields (and field widths) using ASCII art; we extract field names and widths and directly generate data structures (specifically, structs in C) to represent headers to enable automated code generation (\secref{section:codegen}). Some RFCs~\cite{rfc1059} also contain pseudo-code, which we represent as logical forms to facilitate code generation.

\parab{Running CCG.} After pre-processing, we run CCG on each sentence of an RFC. Ideally, CCG should output exactly one logical form for a sentence. In practice, it outputs \textbf{\textit{zero or more}} logical forms, some of which arise from limitations in CCG, and some from ambiguities inherent in the sentence.

\section{Disambiguation}
\label{section:disambiguation}


Next we describe how \sysname leverages domain knowledge to automatically resolve some ambiguities, where semantic parsing resulted in either 0 or more than 1 logical forms.


\subsection{Why Ambiguities Arise}
\label{subsection:why-ambiguities-arise}

To show how we automatically resolve ambiguities, we take examples from the ICMP RFC~\cite{rfc792} for which our semantic parser returned either 0 or more than 1 logical forms.


\parab{Zero logical forms.} Several sentences in the ICMP RFC resulted in zero logical forms after semantic parsing, all of which were grammatically incomplete, lacking a subject:
\begin{description}[itemsep=0mm]
    \item[\textbf{\textit{A}}] \textit{The source network and address from the original datagram's data}
    \item[\textbf{\textit{B}}] \textit{The internet header plus the first 64 bits of the original datagram's data}
    \item[\textbf{\textit{C}}] \textit{If code = 0, identifies the octet where an error was detected}
    \item[\textbf{\textit{D}}] \textit{Address of the gateway to which traffic for the network specified in the internet destination network field of the original datagram's data should be sent}
\end{description}
Such sentences are common in protocol header field descriptions. The last sentence is difficult even for a human to parse.  

\parab{More than 1 logical form.} Several sentences resulted in more than one logical form after semantic parsing.
The following two sentences are \textit{grammatically incorrect}:
\begin{description}[itemsep=0mm]
    \item[\textbf{\textit{E}}] \textit{If code = 0, an identifier to aid in matching timestamp and replies, may be zero}
    \item[\textbf{\textit{F}}] \textit{If code = 0, a sequence number to aid in matching timestamp and replies, may be zero}
\end{description}
The following example needs \textit{additional context}, and contains \textit{imprecise language}:
\begin{description}[itemsep=0mm]
    \item[\textbf{\textit{G}}] \textit{To form a information reply message, the source and destination addresses are simply reversed, the type code changed to 16, and the checksum recomputed}
\end{description}
A machine parser does not realize that source and destination addresses refer to fields in the IP header. Similarly, it is unclear from this sentence whether the checksum refers to the IP checksum or the ICMP checksum. Moreover, the term \textit{type code} is confusing, even to a (lay) human reader, since the ICMP header contains both a \textit{type} field and a \textit{code} field.

Finally, this sentence, discussed earlier (\secref{s:analys-impl}), is under-specified, since it does not describe which byte the checksum computation should \textit{end} at:
\begin{description}[itemsep=0mm]
    \item[\textbf{\textit{H}}] \textit{The checksum is the 16-bit ones's complement of the one's complement sum of the ICMP message starting with the ICMP Type}
\end{description}
While sentences \textbf{\textit{G}} and \textbf{\textit{H}} are grammatically correct and should have resulted in a single logical form, the CCG parser considers them ambiguous as we explain next.


\parab{Causes of ambiguities: zero logical forms.}
Examples \textbf{\textit{A}} through \textbf{\textit{C}} are missing a subject. In the common case when these sentences describe a header field, that header field is usually the subject of the sentence. This information is available to \sysname when it extracts structural information from the RFC (\secref{section:approach}). When a sentence that is part of a field description has zero logical forms, \sysname can re-parse that sentence by supplying the header. This approach does not work for \textbf{\textit{D}}; this is an incomplete sentence, but CCG is unable to parse it even with the supplied header context. Ultimately, we had to re-write that sentence to successfully parse it.

\begin{figure*}
  \scriptsize
  \begin{subfigure}[l]{.5\textwidth}
    \centering
  \begin{tabular}{r|l}
    \tableh{Sentence} & \textit{For computing the checksum, the checksum field should be zero} \\
    \tableh{LF 1} & \predicate{@AdvBefore}(\predicate{@Action}('compute','0'),\predicate{@Is}(\predicate{@And}('checksum\_field','checksum'),'0'))\\
    \tableh{LF 2} & \textbf{\predicate{@AdvBefore}(\predicate{@Action}('compute','checksum'),\predicate{@Is}('checksum\_field','0'))} \\
    \tableh{LF 3} & \predicate{@AdvBefore}('0',\predicate{@Is}(\predicate{@Action}('compute',\predicate{@And}('checksum\_field','checksum')),'0')) \\
    \tableh{LF 4} & \predicate{@AdvBefore}('0',\predicate{@Is}(\predicate{@And}('checksum\_field',\predicate{@Action}('compute','checksum')),'0'))\\
  \end{tabular}
\end{subfigure}
\hspace{6em}
\begin{subfigure}[c]{.3\textwidth}
  \centering
  \resizebox{!}{1.4cm}{\input{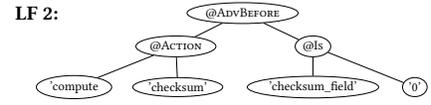}}
\end{subfigure}
\caption{Example of multiple LFs from CCG parsing of ``For computing the checksum, the checksum should be zero''.}
\label{tab:multi-lfs}
\end{figure*}

\parab{Causes of ambiguities: more than one logical form.}
Multiple logical forms arise from more fundamental limitations in machine parsing. Consider \figref{tab:multi-lfs}, which shows multiple logical forms arising for a single sentence. Each logical form consists of \textit{nested predicates} (similar to a statement in a functional language), where each predicate has one or more arguments. A predicate represents a logical relationship (\predicate{@And}), an assignment (\predicate{@Is}), a conditional (\predicate{@If}), or an action (\predicate{@Action}) whose first argument is the name of a function, and subsequent arguments are function parameters. Finally, \figref{tab:multi-lfs} illustrates that a logical form can be naturally represented as a tree, where the internal nodes are predicates and leaves are (scalar) arguments to predicates.

\parae{Inconsistent argument types.}
In some logical forms, their arguments are incorrectly typed, so they are obviously wrong. For example, LF1 in \figref{tab:multi-lfs}, the second argument of the \verb=compute= action must be the name of a function, not a numeric constant. CCG's lexical rules don't support type systems, so cannot eliminate badly-typed logical forms.

\parae{Order-sensitive predicate arguments.}
The parser generates multiple logical forms for the sentence \textbf{\textit{E}}. Among these, in one logical form, \verb=code= is assigned zero, but in the others, the \verb=code= is tested for zero. Sentence \textbf{\textit{E}} has the form ``If A, (then) B'', and CCG generates two different logical forms: \predicate{@If(A,B)} and \predicate{@If(B,A)}. This is not a mistake humans would make, since the condition and action are clear from the sentence. However, CCG's flexibility and expressive power may cause over-generation of semantic interpretations in this circumstance. This unintended behavior is well-known~\cite{hockenmaier-bisk-2010-normal, white2008more}.

\parae{Predicate order-sensitivity.} Consider a sentence of the form ``A of B is C''. In this sentence, CCG generates two distinct logical forms. In one, the \predicate{@Of} predicate is at the root of the tree, in the other \predicate{@Is} is at the root of the tree. The first corresponds to the grouping ``(A of B) is C'' and the second to the grouping ``A of (B is C)''. For sentences of this form, the latter is incorrect, but CCG unable to generate disambiguate between the two.

\parae{Predicate distributivity.} Consider a sentence of the form ``A and B is C''. This sentence exemplifies a grammatical structure called \textit{coordination}~\cite{steedman2011combinatory}\footnote{For example: \emph{Alice sees and Bob says he likes Ice Cream.}}. For such a sentence, CCG will generate two logical forms, corresponding to: ``(A and B) is C'' and ``(A is C) and (B is C)'' (in the latter form, ``C'' distributes over ``A'' and ``B''). In general, both forms are equally correct. However, CCG sometimes chooses to distribute predicates when it should not. This occurs because CCG is unable to distinguish between two uses of the \textit{comma}: one as a conjunction, and the other to separate a dependent clause from an independent clause. In sentences with a comma, CCG generates logical forms for both interpretations. RFCs contain some sentences of the form ``A, B is C''\footnote{If a higher-level protocol uses port numbers, they are assumed to be in the first 64 data bits of the original datagram's data.}. When CCG interprets the comma to mean a conjunction, it generates a logical form corresponding to ``A is C and B is C'', which, for this sentence, is clearly incorrect.

\begin{figure}
    \centering
    \resizebox{\columnwidth}{!}{\input{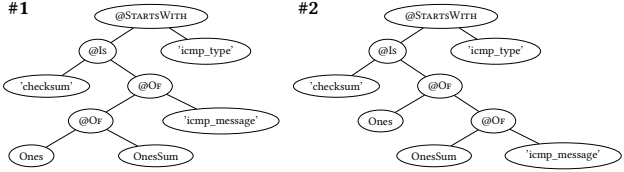}}
    \caption{LF Graphs of sentence \textbf{\textit{H}}.
    }
    \label{fig:lf-graphs}
\end{figure}

\parae{Predicate associativity.}
Consider sentence \textbf{\textit{H}}, which has the form ``A of B of C'', where each of A, B, and C are predicates (\eg A is the predicate \predicate{@Action}("16-bit-ones-complement"). In this example, the CCG parser generates two semantic interpretations corresponding to two different groupings of operations (one that groups A and B, the other that groups B and C: \figref{fig:lf-graphs}). In this case, the \predicate{@Of} predicate is associative, so the two logical forms are equivalent, but the parser does not know this.

\subsection{Winnowing Ambiguous Logical Forms}
\label{s:winn-ambig-logic}

We define the following checks to address each of the above types of ambiguities (\secref{subsection:why-ambiguities-arise}), which \sysname applies to sentences with multiple logical forms, winnowing them down (often) to one logical form (\secref{section:evaluation}). These checks apply broadly because of the restricted way in which specifications use natural language. While we derived these by analyzing ICMP, we show that these checks also help disambiguate text in other RFCs. At the end of this process, if a sentence is still left with multiple logical forms, it is fundamentally ambiguous, so \sysname prompts the user to re-write it.

\parab{Type.} For each predicate, \sysname defines one or more type checks: action predicates have function name arguments, assignments cannot have constants on the left hand side, conditionals must be well-formed, and so on.

\parab{Argument ordering.} For each predicate for which the order of arguments is important, \sysname defines checks that remove logical forms that violate the order.

\parab{Predicate ordering.} For each pair of predicates where one predicate cannot be nested within another, \sysname defines checks that remove order-violating logical forms.

\parab{Distributivity.} To avoid semantic errors due to comma ambiguity, \sysname always selects the non-distributive logical form version (in our example, ``(A and B) is C'').

\parab{Associativity.} If predicates are associative, their logical form trees (\figref{fig:lf-graphs}) will be \textit{isomorphic}. \sysname detects associativity using a standard graph isomorphism algorithm.

\section{Code Generation}
\label{section:codegen}

Next we discuss how we convert the intermediate representation of disambiguated logical forms to code.

\subsection{Challenges}
\label{section:codegen-challenges}

We faced two main challenges in code generation: (a) representing implicit knowledge about dependencies between two protocols or a protocol and the OS and (b) converting a functional logical form into imperative code.

\parab{Encoding protocol and environment dependencies.}
Networked systems rely upon protocol stacks, where protocols higher in the stack use protocols below them. For example, ICMP specifies what operations to perform on IP header fields (\eg sentence \textbf{\textit{G}} in \secref{section:disambiguation}), and does not specify but assumes an implementation of one's complement. Similarly, standards descriptions do not explicitly specify what abstract functionality they require of the underlying operating system (\eg the ability to read interface addresses). 

To address this challenge, \sysname requires a pre-defined \textit{static framework} that provides such functionality along with an API to access and manipulate headers of other protocols, and to interface with the OS. \sysname's generated code (discussed below) uses the static framework. The framework may either contain a complete implementation of the protocols it abstracts, or, more likely, invoke existing implementations of these protocols and services provided by the OS.



\parab{Logical Forms as an Intermediate Representation.}
The parser generates an LF to represent a sentence. For code generation, these sentences (or fragments thereof) fall into two categories: actionable and non-actionable sentences. Actionable sentences result in executable code: they describe value assignments to fields, operations on headers, and computations (\eg checksum). Non-actionable sentences do not specify executable code, but specify a future intent such as \textit{``The checksum may be replaced in the future''} or behavior intended for other protocols such as \textit{``If a higher level protocol uses port numbers, port numbers are assumed to be in the first 64 data bits of the original datagram's data''}. Humans may intervene to identify non-actionable sentences; \sysname tags their logical forms with a special predicate \predicate{@AdvComment}.

The second challenge is that parsers generate logical forms for individual sentences, but the ordering of code generated from these logical forms is not usually explicitly specified. Often the order in which sentences occur matches the order in which to generate code for those sentences. For example, an RFC specifies how to set field values, and it is safe to generate code for these fields in the order in which they appear. There are, however,  exceptions to this. Consider the sentence in \figref{tab:multi-lfs}, which specifies that, when computing the checksum, the checksum field must be zero. This sentence occurs in the RFC \textit{after} the sentence that describes how to compute checksum, but its executable code must occur \textit{before}. To address this, \sysname contains a lexical entry that identifies, and appropriately tags (using a special predicate \predicate{@AdvBefore}), sentences that describe such \textit{advice} (as used in functional and aspect-oriented languages).\footnote{Advice covers statements associated with a function that must be executed before, after, or instead of that function. Here, the checksum must be set to zero \textit{before} computing the checksum.}

\subsection{Logical Forms to Code}
\label{section:codegen-process}

\parab{Pre-processing and contextual information.}
The process of converting logical forms to code is multi-stage, as shown in the right block of \figref{fig:sys_pipeline}. Code generation begins with pre-processing actions. First, \sysname filters out logical forms with the \predicate{@AdvComment} predicate. Then, it prepares logical forms for code conversion by adding contextual information. A logical form does not, by itself, have sufficient information to auto-generate code. For example, from a logical form that says 'Set (message) type to 3' (\predicate{@Is}(type, 3)) it is not clear what ``type'' means and must be inferred from the context in which that sentence occurs. In RFCs, this context is usually implicit from the document structure (the section, paragraph heading, or indentation of text). \sysname auto-generates a \textit{context dictionary} for each logical form (or sentence) to aid code generation (\tabref{t:codegen_example}).

In addition to this dynamic context, \sysname also has a \textit{pre-defined} \textit{static context dictionary} that encapsulates information in the static context. This contains field names used in lower-level protocols (\eg the table maps terms source and destination addresses to corresponding fields in the IP header, or the term ``one's complement sum'' to a function that implements that term). During code generation, \sysname first searches the dynamic context, then the static context. 



\begin{table}
  \centering
  \footnotesize
  \begin{tabular}{r|p{6.5cm}}
    \tableh{LF} & \predicate{@Is}('type', '3')\\
    \tableh{context} & \{"protocol": "ICMP", "message": "Destination Unreachable Message", "field": "type", "role": ""\} \\
    \tableh{code} & \lstinline[style=cpp]!hdr->type = 3;! \\
  \end{tabular}
  \caption{Logical form with context and resulting code.}
  \label{t:codegen_example}
\end{table}

\parab{Code generation.}
After preprocessing, \sysname generates code for a logical form using a post-order traversal of the single logical form obtained after disambiguation. For each predicate, \sysname uses the context to convert the predicate to a code snippet; concatenating these code snippets results in executable code for the logical form.

\sysname then concatenates code for all the logical forms in a message into a packet handling function. In general, for a given message, it is important to distinguish between code executed at the sender versus at the receiver, and to generate two functions, one at the sender and one at the receiver. Whether a logical form applies to the sender or the receiver is also encoded in the context dictionary (\tabref{t:codegen_example}). Also, \sysname uses the context to generate unique names for the function, based on the protocol, the message type, and the role, all of which it obtains from the context dictionaries.

Finally, \sysname processes advice at this stage to decide on the order of the generated executable code. In its current implementation, it only supports \predicate{@AdvBefore}, which inserts code before the invocation of a function. 

\parab{Iterative discovery of non-actionable sentences.} Non-actionable sentences are those for which \sysname should not generate code. Rather than assume that a human annotates each RFC with such sentences before \sysname can execute, \sysname provides support for \textit{iterative discovery} of such sentences, using the observation that \textit{a non-actionable sentence will usually result in a failure during code generation}. So, to discover such sentences, a user runs the RFC through \sysname repeatedly. When it fails to generate code for a sentence, it alerts the user to confirm whether this was a non-actionable sentence or not, and annotates the RFC accordingly. During subsequent passes, it tags the sentence's logical forms with \predicate{@AdvComment}, which the code generator ignores.

In ICMP, for example, there are 35 such sentences. Among RFCs we evaluated, \sysname can automatically tag such code generation failures as \predicate{@AdvComment} without human intervention (\ie there were no cases of an actionable sentence that failed code generation once we defined the context).

\section{Evaluation}
\label{section:evaluation}
Next we quantify \system's ability to find spec ambiguities, its generality across RFCs, and the importance of disambiguation and of our parsing and code generation extensions.

\subsection{Methodology}

\parab{Implementation.}
\system includes a networking dictionary, new CCG-parsable lexicon entries, a set of inconsistency checks, and LF-to-code predicate handler functions. We used the index of \cite{kurosecomputer} to create a dictionary of about 400 terms. \sysname adds 71 lexical entries to an nltk-based CCG parser \cite{loper2002nltk}. Overall, \sysname consists of 7,128 lines of code.


To winnow ambiguous logical forms for ICMP (\secref{s:winn-ambig-logic}), we defined 32 type checks, 7 argument ordering checks, 4 predicate ordering checks, and 1 distributivity check.  Argument ordering and predicate ordering checks maintain a blocklist.  Type checks use an allowlist and are thus the most prevalent.   The distributivity check has a single implicit rule.  For code generation, we defined 25 predicate handler functions to convert LFs to code snippets.

\parab{Test Scenarios.}
First we examine the ICMP RFC, which defines eight ICMP message types.\footnote{ICMP message types include destination unreachable, time exceeded, parameter problem, source quench, redirect, echo/echo reply, timestamp/timestamp reply, and information request/reply.}  Like the student assignments we analyzed earlier, we generated code for each ICMP message type.  
To test this for each message, as with the student projects, the client sends test messages to the router which then responds with the appropriate ICMP message.  For each scenario, we captured both sender and receiver packets and verified correctness with \verb=tcpdump=.  We include details of each scenario in the Appendix. To demonstrate the generality of \system, we also evaluated IGMP, NTP, and BFD.

\subsection{End-to-end Evaluation}
\label{s:end-end-evaluation}


Next we verify that ICMP code generated by \system produces packets that interoperate correctly with Linux tools.  

\parab{Packet capture based verification.} In the first experiment, we examined the packet emitted by a \system-generated ICMP implementation with \verb=tcpdump=~\cite{tcpdump}, to verify that \verb=tcpdump= can read packet contents correctly without warnings or errors. Specifically, for each message type, for both sender and receiver side, we use the static framework in \system-generated code to generate and store the packet in a pcap file and verify it using \verb=tcpdump=. \verb=tcpdump= output lists packet types (\eg an IP packet with a time-exceeded ICMP message) and will warn if a packet of truncated or corrupted packets. In all of our experiments we found that \textit{\textbf{\sysname generated code produces correct packets with no warnings or errors}}.

\parab{Interoperation with existing tools.} Here we test whether a \system-generated ICMP implementation interoperates with tools like \verb=ping= and \verb=traceroute=. To do so, we integrated our static framework code and the \system-generated code into a Mininet-based framework used for the course described in \secref{sec:background}.  With this framework, we verified, with four Linux commands (testing echo, destination unreachable, time exceeded, and traceroute behavior), that a \system-generated receiver or router correctly processes echo request packets sent by \verb=ping= and TTL-limited data packets or packets to non-existent destinations sent by \verb=traceroute=, and its responses are correctly interpreted by those programs. For all these commands, the \textit{\textbf{generated code interoperates correctly with these tools}}.


\subsection{Exploring Generality: IGMP and NTP}
\label{ss:eval-igmp}

To understand the degree to which \system generalizes to other protocols, we ran it on two other protocols: parts of IGMP v1 as specified in RFC 1112~\cite{rfc1112} and NTP~\cite{rfc1059}. In \secref{section:discussion}, we discuss what it will take to extend \sysname to completely parse these RFCs and generalize it to a larger class of protocols.

\parab{IGMP.} In RFC 1112~\cite{rfc1112}, we parsed the packet header description in Appendix I of the RFC. To do this, we added to \sysname 8 lexical entries (beyond the 71 we had added for ICMP entries), 4 predicate function handlers (from 21 for ICMP), and 1 predicate ordering check (from 7 for ICMP). For IGMP, \sysname generates the sending of host membership and query message. 
We also verified  interoperability of the generated code. In our test, our generated code sends a host membership query to a commodity switch. We verified, using packet captures, that the switch's response is correct, indicating that it interoperates with the sender code.


\parab{NTP.} For NTP~\cite{rfc1059}, we parsed Appendices A and B: these describe, respectively, how to encapsulate NTP messages in UDP, and the NTP packet header format and field descriptions. To parse these, we added only 5 additional lexical entries and 1 predicate ordering check beyond what we already had for IGMP and NTP. 

\subsection{Exploring Generality: BFD}
Thus far, we have discussed how \sysname supports headers, field descriptions, constraints, and basic behaviors. We now explore applying \sysname to BFD~\cite{rfc5880}, a recent protocol whose spec contains sentences that describe how to initiate/update state variables. We have use \sysname to parse such \textit{state management} sentences (\S6.8.6 in RFC 5880).


\parab{BFD Introduction.}
BFD is used to detect faults between two nodes. Each node maintains multiple state variables for both protocol and connection state. Connection state is represented by a 3-state machine and represents the status (\eg established, being established, or being torn down) of the session between nodes. Protocol state variables are used to track local and remote configuration.\footnote{This is common across protocols: for example, TCP keeps track of protocol state regarding ACK reception.}

\parab{State Management Dictionary.} 
A state management sentence describes how to use or modify protocol or connection state in terms of state management variables. For example, \textit{bfd.SessionState} is a connection state variable; \textit{Up} is a permitted value. We extend our term dictionary to include these state variables and values as noun phrases.


\begin{table}[h]
  \centering
  \footnotesize
  \begin{tabular}{ccm{6.6cm}}
    \toprule
    \multicolumn{2}{c}{\tableh{Type}} & \tableh{Example} \\
    \midrule
    \multirow{2}{*}{\rotatebox[origin=c]{90}{\textbf{Nested code}}} & \rotatebox[origin=c]{90}{\scriptsize Original} & If the Your Discriminator field is nonzero, it MUST be used to select \emph{\textcolor{blue}{the session}} with which this BFD packet is associated. If \emph{\textcolor{blue}{no session}} is found, the packet MUST be discarded. \\
    \cmidrule{3-3}
    & \rotatebox[origin=c]{90}{\scriptsize Rewritten} & If the Your Discriminator field is nonzero, it MUST be used to select the session with which this BFD packet is associated. If \emph{\textcolor{blue}{the Your Discriminator field is nonzero and}} no session is found, the packet MUST be discarded. \\
    \midrule
    \multirow{2}{*}{\rotatebox[origin=c]{90}{\textbf{Rephrasing}}} & \rotatebox[origin=c]{90}{\scriptsize Original} & If \emph{\textcolor{red}{bfd.RemoteDemandMode is 1}}, bfd.SessionState is Up, and bfd.RemoteSessionState is Up, \emph{\textcolor{red}{Demand mode is active on the remote system}} and the local system MUST cease the periodic transmission of BFD Control packets.\\
    \cmidrule{3-3}
    & \rotatebox[origin=c]{90}{\scriptsize Rewritten} & If bfd.RemoteDemandMode is 1, bfd.SessionState is Up, and bfd.RemoteSessionState is Up, the local system MUST cease the periodic transmission of BFD Control packets.\\
    \bottomrule
  \end{tabular}
  \caption{Challenging BFD state management sentences.}
  \label{t:bfd-new-challenge}
  \vskip -1em
\end{table}


\parab{Limitations and More Challenges.}
When we performed CCG parsing and code generation on state management sentences, we found two types of sentences that could not be parsed correctly (Table~\ref{t:bfd-new-challenge}). Both of these sentences reveal limitations in the underlying NLP approach we use.


The CCG parser treats each sentence independently, but the first example in \tabref{t:bfd-new-challenge} illustrates dependencies across sentences. Specifically, \sysname must infer that the reference to \textit{no session} in the second sentence must be matched to \textit{the session} in the first sentence. This is an instance of the general problem of co-reference resolution~\cite{corecoref}, which can resolve identical noun phrases across sentences. To our knowledge, semantic parsers cannot yet resolve such references. To get \sysname to parse the text, we rewrote the second sentence to clarify the co-reference, as shown in \tabref{t:bfd-new-challenge}.

The second sentence contains three conditionals, followed a non-actionable fragment that rephrases one of the conditionals. Specifically, the first condition \textit{if bfd.RemoteDemandMode is 1}, is rephrased, in English, immediately afterwards (\textit{Demand mode is active on the remote node}). To our knowledge, current NLP techniques cannot easily identify rephrased sentence fragments. \sysname relies on human annotation to identify this fragment as non-actionable; after removing the fragment, it is able to generate code correctly for this sentence.

\parab{Parsing.}
We focus on explaining our analysis of such state management sentences. \sysname is able to parse the packet header as given in RFC 5880\S 4.1. We analyzed 22 state management sentences in RFC 5880\S 6.8.6 which involve a greater diversity of operations than pure packet generation. To support these, we added 15 lexical entries, 10 predicates, and 8 function handlers.

\subsection{Disambiguation}

\begin{table}[t]
  \centering
  \footnotesize
  \begin{tabular}{m{1.25cm}m{4.5cm}c}
    \toprule
    \tableh{Category} & \tableh{Example} & \tableh{Count} \\
    \midrule
    More than 1 LF & To form an echo reply message, the source and destination addresses are simply reversed, the type code changed to 0,
    and the checksum recomputed. & 4 \\
    \midrule
    0 LF & Address of the gateway to which traffic for the network specified in the internet destination network field of the original datagram's data should be sent. & 1 \\
    \midrule
    Imprecise sentence & If code = 0, an identifier to aid in matching echos and replies, may be zero. & 6 \\
    \bottomrule
  \end{tabular}
  \caption{Examples of categorized rewritten text.}
  \label{t:examples-rewrite}
\end{table}

\begin{figure}[t]
    \centering
    \vskip -0.5em
    \resizebox{\columnwidth}{!}{\tikzset{%
  arrow/.style = {
    ->,
  },
  noarrow/.style = {},
  box/.style = {
    draw,
    align = center,
    inner sep = 3pt,
    text centered,
    font = \footnotesize,
    text = black,
    text opacity = 1,
  },
  component/.style = {
    box,
  },
  semanticparsing/.style = {
    fill = cyan,
    fill opacity=.2,
  },
  disambiguation/.style = {
    fill = cyan,
    fill opacity=.2,
  },
  implementation/.style = {
    fill = orange,
    fill opacity=.2,
  },
  decision/.style = {
    box,
    diamond,
    aspect = 2,
    inner sep = 1pt,
  },
}

\def\specicon{
  \begin{scope}[scale=.5]
    \draw [fill=gray, fill opacity=.2]
    (0,0) -- (0, .8) -- (.75, .8) -- (.75, 0.125) -- (.675, 0) -- (0, 0);
    \draw [fill=gray, fill opacity=.4]
    (.75, 0.125) -- (.675, 0.1) -- (.675, 0);
    \node[] at (.375,.575) {{\fontsize{4}{4}\texttt{RFC}}};
    \node[] at (.375,.275) {{\fontsize{4}{4}\texttt{792}}};
  \end{scope}
}

\def\codeicon{
  \begin{scope}[scale=.5]
    \draw [fill=gray, fill opacity=.2]
    (0,0) -- (0, .8) -- (.75, .8) -- (.75, 0.125) -- (.675, 0) -- (0, 0);
    \draw [fill=gray, fill opacity=.4]
    (.75, 0.125) -- (.675, 0.1) -- (.675, 0);
    \node[] at (.375,.575) {{\fontsize{4}{4}\texttt{CODE}}};
  \end{scope}
}

\def\personicon#1{
  \begin{scope}[shift={#1},scale=.5]
  \fill (0,0) arc(0:180:.3 and .6) -- (0, 0);
  \fill (-.3,.75) circle (.2);
  \node[] at (-.3,-.2) {\tiny user};
  \end{scope}
}

\begin{tikzpicture}[>=latex]

  \matrix[column sep = -1] at (-.85, 1.75) {
    \draw[disambiguation, yshift = .5] circle (.1); &&
    \node {{\fontsize{5}{5}\textbf{Disambiguation}}}; \\
  };
  \matrix[column sep = -1] at (.85, 1.75) {
    \draw[implementation, yshift = .5] circle (.1); &&
    \node {{\fontsize{5}{5}\textbf{Implementation}}}; \\
  };

  \matrix[column sep = 120, xshift = -1] at (0, 1.675) {
    \specicon &
    \codeicon \\
  };

  \matrix[column sep = 8] at (0, 1) {
    \node[semanticparsing,component, scale=.5, text width = 32] (SemanticParsing) {Semantic Parsing}; &
    \node[disambiguation,component, scale=.5] (Disambiguation) {Disambiguation}; &
    \node[disambiguation,decision, scale=.5] (SemanticTest) {1 LF/sentence}; &
    \node[implementation, component, scale=.5, text width = 32] (CodeGen) {Code Gen.}; &
    \node[implementation, decision, scale=.5] (UnitTest) {Unit Tests}; \\
  };


  \personicon{(-.5,.15)};


  \draw[arrow] (SemanticParsing.east) -- (Disambiguation.west);
  \draw[arrow] (Disambiguation.east) -- (SemanticTest.west);
  \draw[arrow] (SemanticTest.east) -- (CodeGen.west);
  \draw[arrow] (CodeGen.east) -- (UnitTest.west);

  \node[xshift = 2, yshift = 3] at (SemanticTest.east) {\color{olive}{\fontsize{4}{4}\checkmark}};
  \node[xshift = 2, yshift = -2.5] at (SemanticTest.south) (Xdisamb) {\color{red}{\fontsize{3}{4}\sffamily X}};
  \node[draw, very thin, xshift=11.5, yshift=.5, inner sep=.8, align=left] at (Xdisamb.south) {\fontsize{4}{4}\emph{5 sentences}};

  \draw[arrow] (-2.415, 1.45) -- (SemanticParsing.north);
  \draw[arrow] (UnitTest.north) -- (2.31, 1.45);
  \node[xshift = 4, yshift = 3] at (UnitTest.north) {\color{olive}{\fontsize{4}{4}\checkmark}};
  \node[xshift = 2, yshift = -3] at (UnitTest.south) (Xunit) {\color{red}{\fontsize{3}{4}\sffamily X}};
  \node[draw, very thin, xshift = -15, yshift=-2.5, inner sep=.8, align=left] at (Xunit.south) {\fontsize{4}{4}\emph{6 sentences}};

  \draw[arrow] (SemanticTest.south) |- (-.5 ,.3);
  \draw[noarrow] (-.85, .3) -| (-2.9, 1.4) -| ([yshift=5,xshift=-.05]SemanticParsing.north);
  \draw[noarrow] (UnitTest.south) |- (-.4, .3);
  \node[anchor=south west, yshift = -2.5] at (-2.75, .3)
       {{\fontsize{4}{5}\textit{resolve ambiguity and}}};
  \node[anchor=north west, yshift = 2.5] at (-2.75, .3)
       {{\fontsize{4}{5}\textit{implicit protocol behavior}}};


\end{tikzpicture}

}
    \vskip -0.5em
    \caption{\system workflow in processing RFC 792.}
    \label{fig:sys_workflow}
\end{figure}
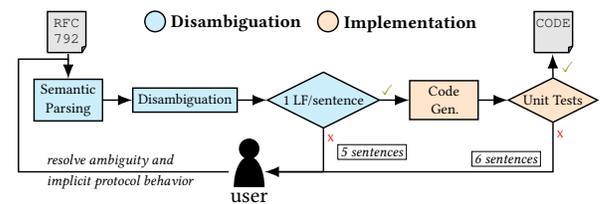

Revising a specification inevitably requires some degree of manual inspection and disambiguation.  \system makes this systematic: it identifies and fixes ambiguities when it can, alerts spec authors or developers when it cannot, and can help iteratively verify re-written parts of the spec.

\parab{Ambiguous sentences.}
When we began to analyze RFC 792 with \system, we immediately found many ambiguities we highlighted throughout this paper; these result in more than one logical form even after manual disambiguation.  

\begin{figure*}[t]
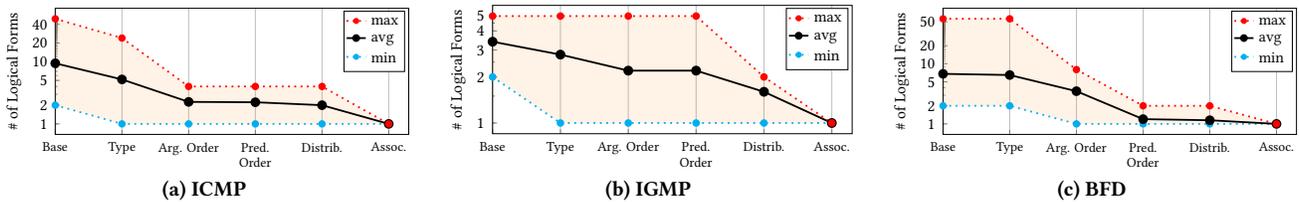

    \captionsetup[subfloat]{skip=1pt}
    \hspace{-0.35cm}
    \subfloat[ICMP]{
        \label{fig:checks-all}
        \resizebox{0.333\textwidth}{!}{\input{figs/checks_common.tex}

\begin{tikzpicture}[scale=1]

  \begin{axis}[
    EchoChecksPlot,
    ylabel style = { font = \scriptsize },
    xlabel style = { font = \footnotesize },
    ymode=log,
    ytick = {1, 2, 5, 10, 20, 40},
    minor ytick = {1, 2, ..., 50},
    log base 10 number format code/.code = {
          $\pgfmathparse{10^(#1)}\pgfmathprintnumber{\pgfmathresult}$
        },
   xmajorgrids={true},
   ylabel = {\# of Logical Forms},
   ]
    \addplot[Min] coordinates {
      (Base,               2)
      (Type,               1)
      (Argument Ordering,  1)
      (Predicate Ordering, 1)
      (Distributivity,     1)
      (Associativity,      1)
    };
    \addplot[Avg] coordinates {
      (Base,               9.43)
      (Type,               5.19)
      (Argument Ordering,  2.26)
      (Predicate Ordering, 2.23)
      (Distributivity,     2)
      (Associativity,      1)
    };
    \addplot[Max] coordinates {
      (Base,              49)
      (Type,              24)
      (Argument Ordering,  4)
      (Predicate Ordering, 4)
      (Distributivity,     4)
      (Associativity,      1)
    };

    \addplot[forget plot, orange, opacity=.1] fill between[of=min and max];

    \addlegendimage{Max};
    \addlegendentry{max};
    \addlegendimage{Avg};
    \addlegendentry{avg};
    \addlegendimage{Min};
    \addlegendentry{min};

  \end{axis}

\end{tikzpicture}

}}\hspace{-0.2cm}%
    \subfloat[IGMP]{
        \label{fig:checks-all-igmp}
        \resizebox{0.333\textwidth}{!}{\input{figs/checks_common.tex}

\begin{tikzpicture}[scale=1]

  \begin{axis}[
    EchoChecksPlot,
    ylabel style = { font = \scriptsize },
    xlabel style = { font = \footnotesize },
    ymode=log,
    ytick = {1, 2, ..., 5},
    minor ytick = {.5, 1, ..., 5},
    log base 10 number format code/.code = {
          $\pgfmathparse{10^(#1)}\pgfmathprintnumber{\pgfmathresult}$
        },
   xmajorgrids={true},
   ylabel = {\# of Logical Forms},
   ]
    \addplot[Min] coordinates {
      (Base,               2)
      (Type,               1)
      (Argument Ordering,  1)
      (Predicate Ordering, 1)
      (Distributivity,     1)
      (Associativity,      1)
    };
    \addplot[Avg] coordinates {
      (Base,               3.4)
      (Type,               2.8)
      (Argument Ordering,  2.2)
      (Predicate Ordering, 2.2)
      (Distributivity,     1.6)
      (Associativity,      1)
    };
    \addplot[Max] coordinates {
      (Base,               5)
      (Type,               5)
      (Argument Ordering,  5)
      (Predicate Ordering, 5)
      (Distributivity,     2)
      (Associativity,      1)
    };

    \addplot[forget plot, orange, opacity=.1] fill between[of=min and max];

    \addlegendimage{Max};
    \addlegendentry{max};
    \addlegendimage{Avg};
    \addlegendentry{avg};
    \addlegendimage{Min};
    \addlegendentry{min};

  \end{axis}

\end{tikzpicture}

}}\hspace{-0.2cm}%
    \subfloat[BFD]{
        \label{fig:checks-all-bfd} 
        \resizebox{0.333\textwidth}{!}{\input{figs/checks_common.tex}

\begin{tikzpicture}[scale=1]

  \begin{axis}[
    EchoChecksPlot,
    ylabel style = { font = \scriptsize },
    xlabel style = { font = \footnotesize },
    ymode=log,
    ytick = {1, 2, 5, 10, 20, 50},
    minor ytick = {1, 2, ..., 56},
    log base 10 number format code/.code = {
          $\pgfmathparse{10^(#1)}\pgfmathprintnumber{\pgfmathresult}$
        },
   xmajorgrids={true},
   ylabel = {\# of Logical Forms},
   ]
    \addplot[Min] coordinates {
      (Base,               2)
      (Type,               2)
      (Argument Ordering,  1)
      (Predicate Ordering, 1)
      (Distributivity,     1)
      (Associativity,      1)
    };
    \addplot[Avg] coordinates {
      (Base,               6.8)
      (Type,               6.5)
      (Argument Ordering,  3.5)
      (Predicate Ordering, 1.2)
      (Distributivity,     1.15)
      (Associativity,      1)
    };
    \addplot[Max] coordinates {
      (Base,               56)
      (Type,               56)
      (Argument Ordering,  8)
      (Predicate Ordering, 2)
      (Distributivity,     2)
      (Associativity,      1)
    };

    \addplot[forget plot, orange, opacity=.1] fill between[of=min and max];

    \addlegendimage{Max};
    \addlegendentry{max};
    \addlegendimage{Avg};
    \addlegendentry{avg};
    \addlegendimage{Min};
    \addlegendentry{min};

  \end{axis}

\end{tikzpicture}

}}\\
    \vskip -.5em
    \caption{Number of LFs after Inconsistency Checks on ICMP/IGMP/BFD text: for each ambiguous sentence, sequentially executing checks on LFs (Base) reduces inconsistencies; after the last Associativity check, the final output is a single LF.}
    \label{fig:disambiguation-checks}
\end{figure*}

%
%

We also encountered ostensibly disambiguated text that yields zero logical forms; this is caused by incomplete sentences.  For example, ``If code = 0, identifies the octet where an error was detected'' is an example that fails CCG parsing due to lack of subject in the sentence, and indeed it may not be parseable for a human lacking context regarding the referent.  Such sentence fragments require human guesswork, but, as we have observed in \secref{section:disambiguation}, we can leverage structural context in the RFC in cases where the referent of these sentences is a field name. In these cases, \sysname is able to correctly parse the sentence by supplying the parser with the subject.

Among 87 instances in RFC 792, we found 4 that result in more than 1 logical form and 1 results in 0 logical forms (Table~\ref{t:examples-rewrite}). We rewrote these 5 ambiguous (of which only 3 are unique) sentences to enable automated protocol generation. These ambiguous sentences were found after \system had applied its checks (\secref{s:winn-ambig-logic})---these are in a sense true ambiguities in the ICMP RFC. In \sysname, we require the user to revise such sentences, according to the feedback loop as shown in Figure~\ref{fig:sys_workflow}. As guidance for future users of \sysname,  the resulting LFs from an ambiguous sentence are all kept after the disambiguation checks are applied and comparing these LFs can guide the users where the ambiguity lies, thus guiding their revisions. In our end-to-end experiments (\secref{s:end-end-evaluation}), we evaluated \sysname using the modified RFC with these ambiguities fixed. 

\parab{Under-specified behavior.}
\sysname can also discover under-specified behavior through unit testing; generated code can be applied to unit tests to see if the protocol implementation is complete. In this process, we discovered 6 sentences that are variants of this sentence:
``If code = 0, an identifier to aid in matching echos and replies, may be zero''. This sentence does not specify whether the sender or the receiver or both can (potentially) set the identifier. The correct behavior is only for the sender to follow this instruction; a sender may generate a non-zero identifier, and the receiver should set the identifier to be zero in the reply. Not doing so results in a non-interoperability with Linux's \textit{ping} implementation.


\begin{figure}[t]
  \centering
  \resizebox{8.3cm}{!}{\input{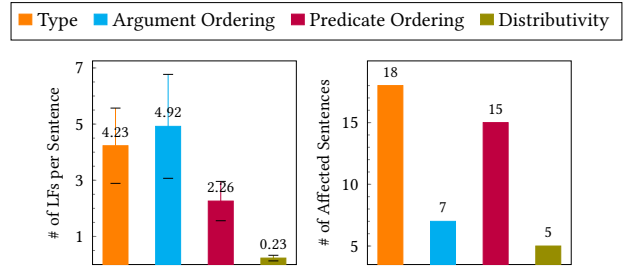}}
  \vskip -.5em
  \caption{Effect of individual disambiguation checks on RFC 792: \textit{Left:} average number of LFs filtered by the check per ambiguous sentence with standard error \textit{Right:} number of ambiguous sentences affected out of 42 total.}
  \label{fig:check-eval}
\end{figure}


\parab{Efficacy of logical form winnowing.}
\sysname winnows logical forms so it can automatically disambiguate text when possible, thereby reducing manual labor in disambiguation. To understand why winnowing is necessary, and how effective each of its checks can be, we collect text fragments that could lead to multiple logical forms, and calculate how many are generated before and after we perform inconsistency checks along with the isomorphism check. We show the extent to which each check is effective in reducing logical forms: in \figref{fig:checks-all}, the max line shows the description that leads to the highest count of generated logical forms and shows how the value goes down to one after all checks are completed.  Similarly, the min line represents the situation for the text that generates the fewest logical forms before applying checks.  Between the min and max lines, we also show the average trend among all sentences.

\figref{fig:checks-all} shows that all sentences resulted in 2-46 LFs, but \sysname's winnowing reduces this to 1 (after human-in-the-loop rewriting of true ambiguities). Of these, type, argument ordering and the associativity checks are the most effective. We apply the same analysis to IGMP (\figref{fig:checks-all-igmp}). In IGMP, the distributivity check is also important. This analysis shows the cumulative effect of applying checks in the order shown in the figure.
We also apply the same analysis to BFD state management sentences (\figref{fig:checks-all-bfd}). We discover some longer sentences could result in up to 56 LFs.

\begin{table}[t]
  \centering
  \footnotesize
  \begin{tabular}{m{5.8cm}cc}
    \toprule
    \tableh{Sentence} & \tableh{Label} & \tableh{\#LFs} \\
    \midrule
    The 'address' of the 'source' in an 'echo message' will be the 'destination' of the \textcolor{red}{'echo reply' 'message'}. & Poor & 16 \\
    \midrule
    The 'address' of the 'source' in an 'echo message' will be the 'destination' of the \textcolor{olive}{'echo reply message'}. & Good & 6 \\
    \bottomrule
  \end{tabular}
  \caption{Comparison of the number of logical forms (LFs) between good and poor noun phrase labels.}
  \label{t:np-lf}
  \vskip -1em
\end{table}

A more direct way to understand the efficacy of checks is shown in \figref{fig:check-eval} (for ICMP). To generate this figure, for each sentence, we apply only one check on the base set of logical forms and measure how many LFs the check can reduce. The graphs show the mean and standard deviation of this number across sentences, and the number of sentences to which a check applies. For ICMP, as before, type and predicate ordering checks reduced LFs for the most number of sentences, but argument ordering reduced the most logical forms. For IGMP (omitted for brevity), the distributivity checks were also effective, reducing one LF every 2 sentences.




\begin{table}[t]
    \scriptsize
    \begin{tabular}{lrrr}
    \toprule
    & \tableh{Increase} & \tableh{Decrease} & \tableh{Zero}\\
    \midrule
    Domain-specific Dict. & 17 & 0 & 0\\
    Noun-phrase Labeling & 0 & 8 & 54\\
    \bottomrule
    \end{tabular}
    \caption{Effect of disabling domain-specific dictionary and noun-phrase labeling on number of logical forms.}
    \label{t:np-comparison}
\end{table}

\parab{Importance of Noun Phrase Labeling.}
\sysname requires careful labeling of noun-phrases using SpaCy based on a domain-specific dictionary (\secref{section:approach}). This is an important step that can significantly reduce the number of LFs for a sentence. To understand why, consider the example in \tabref{t:np-lf}, which shows two different noun-phrase labels, which differ in the way \sysname labels the fragment ``echo reply message''. When the entire fragment is not labeled as a single noun phrase, CCG outputs many more logical forms, making it harder to disambiguate the sentence. In the limit, when \sysname does not use careful noun phrase labeling, CCG is unable to parse some sentences at all (resulting in 0 LFs).

\tabref{t:np-comparison} quantifies the importance of these components. Removing the domain-specific dictionary increases the number of logical forms (before winnowing) for 17 of the 87 sentences in the ICMP RFC. Completely removing noun-phrase labeling using SpaCy has more serious consequences: 54 sentences result in 0 LF. Eight other sentences result in fewer LFs, but these reduce to zero after winnowing.

\section{\sysname Extensions}
\label{section:discussion}

\begin{table}[t]
  \centering
  \resizebox{\columnwidth}{!}{%
    \begin{tabular}{rccccccccc}
      & \rot{\tableh{IPv4}} & \rot{\tableh{TCP}}  & \rot{\tableh{UDP}}  & \rot{\tableh{ICMP}} & \rot{\tableh{NTP}}   & \rot{\tableh{OSPF2}} & \rot{\tableh{BGP4}} & \rot{\tableh{RTP}} & \rot{\tableh{BFD}} \\
      \midrule
      \textbf{\fullmark Packet Format}            & \OK & \OK & \OK & \OK & \OK & \OK & \OK & \OK & \OK \\
      \textbf{\fullmark Interoperation}           & \OK & \OK & \OK & \OK & \OK & \OK & \OK & \OK & \\
      \textbf{\fullmark Pseudo Code}              & \OK & \OK & \OK & \OK & \OK & \OK & \OK & \OK & \OK \\
      \textbf{\partialmark State/Session Mngmt.} &     & \OK &     &     &     & \OK & \OK & \OK & \OK \\
      \textbf{Comm. Patterns}   & \OK & \OK &     &     &     & \OK & \OK & \OK & \OK \\
      \textbf{Architecture}             &     &     &     &     & \OK & \OK &     & \OK &  \\
      \bottomrule
  \end{tabular}
  }
  \caption{Conceptual components in RFCs. \sysname supports components marked with \fullmark (fully) and \partialmark (partially).}
  \label{t:rfc-concepts}
  \vskip -1em  
\end{table}

\begin{table}[t]
  \centering
  \resizebox{\columnwidth}{!}{%
    \begin{tabular}{rccccccccc}
      & \rot{\tableh{IPv4}} & \rot{\tableh{TCP}}  & \rot{\tableh{UDP}}  & \rot{\tableh{ICMP}} & \rot{\tableh{NTP}}   & \rot{\tableh{OSPF2}} & \rot{\tableh{BGP4}} & \rot{\tableh{RTP}} & \rot{\tableh{BFD}} \\
      \midrule
      \textbf{\fullmark Header Diagram}  & \OK & \OK & \OK & \OK & \OK & \OK & \OK & \OK& \OK \\
      \textbf{\fullmark Listing}                & \OK & \OK & \OK & \OK & \OK & \OK & \OK & \OK & \OK \\
      \textbf{Table}                  & \OK & \OK &     &     & \OK & \OK & \OK & \OK & \OK \\
      \textbf{Algorithm Description}  & \OK & \OK &     &     & \OK & \OK &     & \OK & \OK \\
      \textbf{Other Figures}          & \OK &     &     &     & \OK & \OK & \OK & \OK &  \\
      \textbf{Seq./Comm. Diagram}     & \OK & \OK &     &     & \OK & \OK &     & \OK &  \\
      \textbf{State Machine Diagram}  &     & \OK &     &     &     &     &     &     & \OK \\
      \bottomrule
  \end{tabular}
  }
  \caption{Syntactic components in RFCs. \sysname supports parsing the syntax of those marked with \fullmark (fully) and \partialmark (partially).}
  \label{t:rfc-elements}
  \vskip -.75em
\end{table}

\begin{table}[t]
  \centering
  \footnotesize
  \begin{tabular}{r|p{6.1cm}}
    \tableh{sentence} & The timeout procedure is called in client mode and symmetric mode when the peer timer reaches the value of the timer threshold variable. \\
    \tableh{} & \\[-1em]
    \tableh{code} & %
\begin{lstlisting}[style=cpp, belowskip = -1.5em, aboveskip = -.75em]
if (peer.timer >= peer.threshold) {
  if (symmetric_mode || client_mode) {
    timeout_procedure();
  }
}
\end{lstlisting}\\
  \end{tabular}
  \caption{NTP peer variable sentence and resulting code.}
  \label{t:peer-variable}
\end{table}

While \sysname takes a big step toward automated spec processing, much work (likely several papers worth!) remains.

\parab{Specification components.}
To understand this gap, we have manually inspected several protocol specifications and categorized components of specifications into two categories: syntactic and conceptual. Conceptual components (\tabref{t:rfc-concepts})  describe protocol structure and behavior: these include header field semantic descriptions, specification of sender and receiver behavior, who should communicate with whom, how sessions should be managed, and how protocol implementations should be architected. Most popular standards have many, if not all, of these elements. \sysname supports parsing of 3 of the 6 elements in the table, for ICMP and parts of NTP. Our results (\secref{s:end-end-evaluation}) show that extending these elements to other protocols will require marginal extensions at each step. However, much work remains to achieve complete generality, of which state and session management is a significant piece.

\sysname is already able to parse state management. As an example, the NTP RFC has complex sentences on maintaining peer and system variables, to decide when each procedure should be called and when variables should be updated. One example sentence, shown in Table~\ref{t:peer-variable}, concerns when to trigger timeout. \sysname is able to parse the sentence into an LF and turn it into a code snippet. However, NTP requires more complex co-reference resolution, as other protocols may too~\cite{allencoref,corecoref}: in NTP, context for state management is spread throughout the RFC and \sysname will need to associate these conceptual references. For instance, the word ``and'' in the example (Table~\ref{t:peer-variable}) could be equivalent to a logical AND \emph{or} a logical OR operator depending on whether symmetric mode and client mode are mutually exclusive or not. A separate section clarifies that the correct semantics is OR.

RFC authors augment conceptual text with syntactic components (\tabref{t:rfc-elements}). These include forms that provide better understanding of a given idea (\eg header diagrams, tables, state machine descriptions, communication diagrams, and algorithm descriptions). \sysname includes support for two of these elements; adding support for others is not conceptually difficult, but may require significant programming effort.
While much work remains, two significant protocols may be within reach with the addition of complex state management and state machine diagrams: TCP and BGP. 


\parab{Toward full automation?}
Ideally it would be possible to fully automate translation of natural-language specifications to code.  Alas, we believe that the inherent complexity of natural language combined with the inherent logical complexities of protocols and programming languages make it unlikely that this will ever be fully realizable.  However, we believe that it is possible to come close to this, and have aimed to build \system as a first big step in this direction.

The key challenge would be to minimize the manual labor required to assist in disambiguation. Our winnowing already does this (\secref{s:winn-ambig-logic}), but future work will need to explore good user interfaces for human input when \sysname generates 0 LFs or more than 1 LF (\figref{fig:sys_workflow}). \sysname will also need to develop ways for humans to specify cross-references (references to other protocols in a spec), and to write unit tests.

\section{Related Work}
\label{s:related-work}

\parab{Protocol Languages / Formal Specification Techniques.}
Numerous protocol languages have been proposed over the years. In the '80s, Estelle \cite{budkowski1987introduction} and LOTOS \cite{bolognesi1987introduction} provided formal descriptions for OSI protocol suites. Although these formal techniques can specify precise protocol behavior, it is hard for people to understand and thus use for specification or implementation. Estelle used finite state machine specs to depict how protocols communicate in parallel, passing on complexity, unreadability, and rigidity to followup work \cite{boussinot1991esterel,sidhu1990formal,von1987methods}.
Other research such as RTAG~\cite{anderson1988automated}, x-kernel~\cite{hutchinson1991x}, Morpheus~\cite{abbott1993language}, Prolac~\cite{kohler1999readable}, Network Packet Representation~\cite{10.1145/3404868.3406671}, and NCT~\cite{mcmillan2019formal} gradually improved readability, structure, and performance of protocols, spanning specification, testing, and implementation. However, we find and the networking community has found through experience, that English-language specifications are more readable than such protocol languages.

\parab{Protocol Analysis.} Past research \cite{bhargavan2002formal,bishop2005rigorous, bolognesi1987introduction} developed techniques to reason about protocol behaviors in an effort to minimize bugs. Such techniques used finite state machines, higher-order logic, or domain-specific languages to verify protocols. Another thread of work \cite{killian2007life, killian2007mace, lee2012gatling} explored the use of explicit-state model-checkers to find bugs in protocol implementations. This thread also inspired work (\eg \cite{pedrosa2015analyzing}) on discovering non-interoperabilities in protocol implementations.
While our aims are similar, our focus is end-to-end, from specification to implementation, and on identifying where specification ambiguity leads to bugs.

\parab{NLP for Log Mining and Parsing.}
Log mining and parsing are techniques that leverage log files to discover and classify different system events (e.g., 'information', 'warning', and 'error').
Past studies have explored Principal Component Analysis \cite{xu2009detecting}, rule-based analysis \cite{fu2009execution}, statistic analysis \cite{nagaraj2012structured, vaarandi2003data}, and ML-based methods \cite{sipos2014log} to solve log analysis problems. Recent work \cite{aussel2018improving, bertero2017experience} has applied NLP to extract semantic meanings from log files for event categorization.
\sysname is complementary to this line of work.

\parab{Program Synthesis.} To automatically generate code, prior work has explored program synthesis. Reactive synthesis \cite{pnueli1989synthesis, piterman2006synthesis} relies on interaction with users to read input for generating output programs. Inductive synthesis \cite{alur2014syntax} recursively learns logic or functions with incomplete specifications. Proof-based synthesis (\eg \cite{srivastava2010program}) takes a correct-by-construction approach to develop inductive proofs to extract programs. Type-based synthesis \cite{osera2015type, feser2015synthesizing} takes advantage of the types provided in specifications to refine output. In networking, program synthesis techniques can automate (\eg \cite{mcclurg2015efficient, mcclurg2016event}) updating of network configurations, and generating programmable switch code~\cite{10.1145/3387514.3405852}. It may be possible to use program synthesis in \sysname to generate protocol fragments.
\parab{Semantic Parsing and Code Generation.}
Semantic parsing is a fundamental task in NLP that aims to transform unstructured text into structured LFs for subsequent execution~\cite{berant2013semantic}. For example, to answer the question ``Which team does Frank Hoffman play for?'', a semantic parser generates a structured query ``SELECT TEAM from table where PLAYER=Frank Hoffman'' with SQL Standard Grammar~\cite{Date1987AGT}. A SQL interpreter can execute this query on a database and give the correct answer~\cite{kamath2018survey}. Apart from the application to question answering, semantic parsing has also been successful in navigating robots~\cite{tellex2011understanding}, understanding instructions \cite{chen2011learning}, and playing language games~\cite{wang2016learning}.
Research in generating code from natural language goes beyond LFs, to output concrete implementations in high-level general-purpose programming languages~\cite{ling2016latent}. This problem is usually formulated as syntax-constrained sequence generation \cite{yin-neubig-2017-syntactic, liang2016neural}. The two topics are closely related to our work since the process of implementing network protocols from RFCs requires the ability to understand and execute instructions.

\parab{Pre-trained Language Models.} Recently, high-capacity pre-trained language models \cite{peters2018deep, devlin2018bert, yang2019xlnet, lan2019albert} have dramatically improved NLP in question answering, natural language inference, text classification, \etc The general approach is to first train a model on a huge corpus with unsupervised learning (\ie pre-training), then re-use these weights to initialize a task-specific model that is later trained with labeled data (\ie, fine-tuning). In the context of \sysname, such pre-trained models advance improve semantic parsing \cite{Zhang2019BroadCoverageSP, Zhang2019AMRPA}. Recent work \cite{Feng2020CodeBERTAP} also attempts to pre-train on programming and natural languages simultaneously, and achieves state-of-the-art performance in code search and code documentation generation. However, direct code generation using pre-trained language models is an open research area and requires massive datasets; the best model for a related problem, natural language generation, GPT~\cite{radford2019language}, requires 8~M web pages for training.

\section{Conclusions}
\label{s:conclusion}

This paper describes \system, which introduces semi-automated protocol processing across multiple protocol specifications. \sysname includes domain-specific extensions to semantic parsing and automated discovery of ambiguities and enables disambiguation; \system can convert these specs to code.
Future work can extend \sysname to parse more spec elements, and devise better methods to involve humans in the loop to detect and fix ambiguities and guide the search for bugs.\\[1ex]
(This work does not raise any ethical issues.)

{\small \bibliographystyle{acm}
\bibliography{references.bib}}

\pagebreak

\begin{figure*}[t]
    \centering
    \includegraphics[trim=0cm 11.5cm 0.5cm 0cm,clip=true]{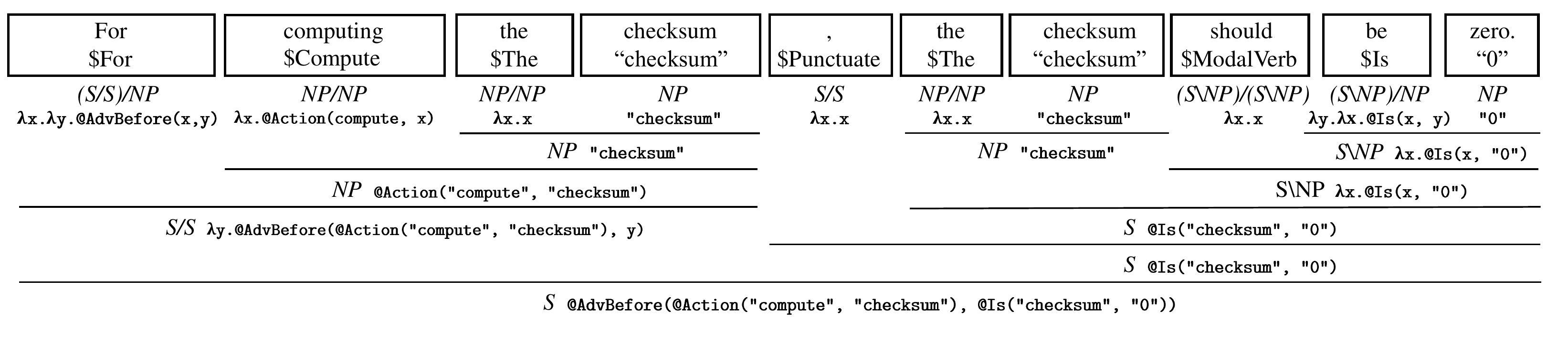}
    \vskip -1em
    \caption{Constructing one logical form from: \textit{``For computing the checksum, the checksum should be zero'' with CCG.}}
    \label{fig:ccg_parsing}
    \vskip -1em
\end{figure*}
\section*{Appendix}
\label{sec:appendix}

\renewcommand{\thesubsection}{\Alph{subsection}}
\setcounter{subsection}{0}

\subsection{ICMP Test Scenario Setup}
\parab{Destination Unreachable Message.}
At the router/receiver side, we assume the router only recognizes three subnets, which are 10.0.1.1/24, 192.168.2.1/24, and 172.64.3.1/24.
At the sender side, we craft the packet with destination IP address not belonging to any of the three subnets.
The receiver reads the packet and calls the generated function to construct the destination unreachable message back to the sender.

\parab{Time Exceeded Message.}
At the sender side, we intentionally generate a packet with the time-to-live field in IP header set to 1, and the destination IP address set to server 1's address.
At the router side, the router checks  the value of time-to-live field and recognizes the packet cannot reach the destination before the time-to-live field counts down to zero.
The router interface calls the generated function to construct a time exceed message and sends it back to the client.

\parab{Parameter Problem Message.}
At the router side, we assume the router can only handle IP packets in which the type of service value equals to zero.
At the sender side, we modify the sent packet to set the type of service value to one.
The router interface recognizes the unsupported type of service value and calls the generated function to construct a parameter problem message back to the client.

\parab{Source Quench Message.}
At the receiver side, we assume one outbound buffer is full, and therefore there is no space to hold a new datagram.
At the sender side, we generated a packet to server 1.
If there is still buffer space for the router to forward the packet to server 1, the router should push the packet to the outbound buffer connected to the subnet where server 1 belongs to.
Under this scenario, the router will decide to discard the received packet, and construct a source quench packet back to the client.

\parab{Redirect Message.}
At the sender side, the client generated a packet to an IP address that is within the same subnet, but sent to the router.
The router discovered the next gateway is in the same subnet as the sender host, and therefore constructs the redirect message to the client with the redirect gateway address by calling the generated functions.

\parab{Echo and Echo Reply Message.}
In RFC 792, echo/echo reply are explained together, but some sentences are merely for echo  while some are only for echo reply.
After analysis, \system generates two different pieces of code.
One is specific to the sender side, and the other is specific to the receiver side.
The client calls the generated function to construct an echo message to the router interface.
The router interface finds it is the destination and constructs an echo reply message back to the client by calling the receiver code.

\parab{Timestamp and Timestamp Reply Message.}
The sender and receiver behavior in this scenario is identical to echo/echo reply.
The sender sends a packet by calling the generated function and the receiver matches the ICMP type and replies to packets with the generated function.
The difference lies in the packet generated by the function.
The timestamp or timestamp reply message do not have datagram data, but they have three different timestamp fields in its header.
The generated function correctly separates three different timestamps with respect to the roles and computation time.

\parab{Information Request and Reply Message.}
The sender and receiver behavior of this scenario is the same as echo/echo reply and timestamp/timestamp reply.
Similar to timestamp/timestamp reply, the differences lie in the generated packets that do not have data; the field values are different.

\subsection{CCG Parsing Example}
We show a more complex example, of deriving one final logical form from the sentence: ``For computing the checksum, the checksum should be zero.'' in \figref{fig:ccg_parsing}. First, each word in the sentence is mapped to its lexical entries (\textit{e.g.}, checksum $\rightarrow$ {NP: "checksum"}). Multiple lexical entries may be available for one word; in this case we make multiple attempts to parse the whole sentence with each entry. After this step, the CCG parsing algorithm automatically applies combination rules and derives final logical forms for the whole sentence.

\end{document}